\documentclass[a4paper]{article}

\usepackage{INTERSPEECH2019}
\usepackage{cite}
\usepackage{microtype}
\usepackage{booktabs}
\usepackage{enumitem}
\usepackage{footnote}
\usepackage{balance}
\makesavenoteenv{tabular}
\usepackage[usenames,dvipsnames]{color}
\usepackage{breakurl}
\usepackage[colorlinks=true]{hyperref}
\hypersetup{
  linkcolor      = {RoyalBlue},
  citecolor      = {RoyalBlue},
  urlcolor       = {RoyalBlue}}

\title{CHiME-6 Challenge:\\Tackling Multispeaker Speech Recognition for Unsegmented Recordings}
\name{$^1$Shinji Watanabe, $^2$Michael Mandel, $^3$Jon Barker, $^4$Emmanuel Vincent\\
$^1$Ashish Arora, $^1$Xuankai Chang, $^1$Sanjeev Khudanpur, $^1$Vimal Manohar, $^1$Daniel Povey, $^1$Desh Raj, $^1$David Snyder, $^1$Aswin Shanmugam Subramanian, $^1$Jan Trmal, $^1$Bar Ben Yair, $^5$Christoph Boeddeker, $^2$Zhaoheng Ni, $^6$Yusuke Fujita, $^6$Shota Horiguchi, $^7$Naoyuki Kanda, $^7$Takuya Yoshioka, $^8$Neville Ryant
}
\address{
  $^1$Johns Hopkins University, USA, $^2$The City University of New York, USA, $^3$University of Sheffield, UK, $^4$Inria, France,  $^5$Paderborn University, Germany, $^6$Hitachi, Ltd., Japan, $^7$Microsoft, USA, $^8$Linguistic Data Consortium, USA}
\email{shinjiw@ieee.org}

\begin{document}

\maketitle
\begin{abstract}
Following the success of the 1st, 2nd, 3rd, 4th and 5th CHiME challenges we organize the 6th CHiME Speech Separation and Recognition Challenge (CHiME-6). 
The new challenge revisits the previous CHiME-5 challenge and further considers the problem of distant multi-microphone conversational speech diarization and recognition in everyday home environments. 
Speech material is the same as the previous CHiME-5 recordings except for accurate array synchronization.
The material was elicited using a dinner party scenario with efforts taken to capture data that is representative of natural conversational speech.
This paper provides a baseline description of the CHiME-6 challenge for both segmented multispeaker speech recognition (Track 1) and unsegmented multispeaker speech recognition (Track 2).
Of note, Track 2 is the first challenge activity in the community to tackle an unsegmented multispeaker speech recognition scenario with a complete set of reproducible open source baselines providing speech enhancement, speaker diarization, and speech recognition modules.
\end{abstract}

\section{Introduction}
\label{sec:intro}
Automatic speech recognition (ASR) performance in difficult reverberant and noisy conditions has improved tremendously in the last decade \cite{VirtanenTechniquesNoiseRobustness2012, Li2015, newera, Vincent2018, Makino2018, haeb2019speech}. This can be attributed to advances in speech processing, audio enhancement, and machine learning, but also to the availability of real speech corpora recorded in cars \cite{SpeechDat-_Car_1, CUMove}, quiet indoor environments \cite{TED, MC-WSJ-AV_PASCAL_SSC2_2012_MMA_REVERB_RealData_2}, noisy indoor and outdoor environments \cite{CHiME3, CHiME4}, and challenging broadcast media \cite{ETAPE, MGB}. Among the applications of robust ASR, voice command in domestic environments has attracted a great deal of interest recently, due in particular to the release of the Amazon Echo, Google Home and other devices targeting home automation and multimedia systems. The CHiME-1 \cite{Barker2013} and CHiME-2 \cite{CHiME_1_CHiME_2_Grid_1} challenges and corpora have contributed to popularizing research on this topic, together with the DICIT \cite{DICIT_1}, Sweet-Home \cite{SweetHome}, and DIRHA \cite{Ravanelli2015} corpora. These corpora feature single-speaker reverberant and/or noisy speech recorded or simulated in a single home, which precludes the use of modern speech enhancement techniques based on machine learning. The two voiceHome corpora \cite{voiceHome,voiceHome2} address this issue, but they are fairly small.

In parallel to research on acoustic robustness, research on conversational speech recognition has also made great progress, as illustrated by the recent announcements of super-human performance \cite{Xiong2017, Saon2017} achieved on the Switchboard telephone conversation task \cite{Switchboard} and by the ASpIRE challenge \cite{harper2015automatic}. Distant-microphone recognition of noisy, overlapping, conversational speech is now widely believed to be the next frontier. Early attempts in this direction can be traced back to the ICSI \cite{ICSI}, CHIL \cite{CHIL}, and AMI \cite{AMI} meeting corpora, the LLSEC \cite{LLSEC} and COSINE \cite{COSINE} face-to-face interaction corpora, and the Sheffield Wargames corpus \cite{SWC}. These corpora were recorded using advanced microphone array prototypes which are not commercially available, and as result could only be installed in a few laboratory rooms. The VOiCES corpus \cite{richey2018voices} utilizes an ad-hoc array of commercial microphones, with pre-recorded speech and noise played over speakers. The DIPCO corpus \cite{van2019dipco}, inspired by the CHiME-5 challenge \cite{Fifth-Barker2018} but of shorter duration, provides recordings of dinner table interactions between four participants recorded simultaneously on several commercially available microphone arrays.  The Santa Barbara Corpus of Spoken American English [30] stands out as the only large-scale corpus of naturally occurring spoken interactions between a wide variety of people recorded in real everyday situations including face-to-face or telephone conversations, card games, food preparation, on-the-job talk, story-telling, and more. Unfortunately, it was recorded via a single microphone. 

The CHiME-6 challenge, which builds upon CHiME-5 \cite{Fifth-Barker2018}, targets the problem of distant microphone conversational speech
recognition in everyday home environments. The speech material has been
collected from twenty real dinner parties that have taken place in real
homes. The recordings have been made using multiple commercially available 4-channel microphone
arrays and have been fully transcribed. The challenge features:
\begin{itemize}[noitemsep]
\item simultaneous recordings from multiple microphone arrays;
\item real conversation, i.e.~talkers speaking in a relaxed and unscripted fashion;
\item a range of room acoustics from 20 different homes each with two or three separate recording areas;
\item real domestic background noises, e.g., kitchen appliances, air conditioning, movement, etc.
\end{itemize}

In the following, we introduce the recording scenario and the two proposed challenge tracks in Section \ref{sec:scenario}. We describe the software baselines and the challenge instructions for Tracks 1 and 2 in Sections \ref{sec:track1_software} and \ref{sec:track2_software}, respectively. We report the corresponding results in Section \ref{sec:results} and conclude in Section \ref{sec:summary}. More details can be found on the challenge website: \\

\centerline{\url{https://chimechallenge.github.io/chime6/}}

\section{Scenario and tracks}
\label{sec:scenario}
\subsection{The scenario}\label{the-scenario}
The dataset is made up of the recording of twenty separate dinner parties taking place in real homes. Each dinner party has four participants --- two acting as hosts and two as guests. 
The party members are all friends who know each other well and who are instructed to behave naturally. 
Efforts have been taken to make the parties as natural as possible. The only constraints are that each party should last a minimum of 2 hours and should be composed of three phases, each
corresponding to a different location:
\begin{itemize}
\item kitchen: preparing the meal in the kitchen area
\item dining: eating the meal in the dining area
\item living: a post-dinner period in a separate living room area
\end{itemize}
Participants have been allowed to move naturally from one location to another but with the instruction that each phase should last at least 30 minutes. 
Participants are free to converse on topics of their choosing --- there is no artificial scenario. 
Some personally identifying material has been redacted post-recording as part of the consent process. Background television and commercial music has been disallowed in order to avoid capturing copyrighted content.

\subsection{The recording set up}\label{the-recording-set-up}
Each party has been recorded with a set of six Microsoft Kinect devices.
The devices have been strategically placed such that there are always at least two capturing the activity in each location. 
Each Kinect device has a linear array of 4 sample-synchronised microphones and a camera.
The raw microphone signals and video have been recorded. 
Each Kinect is recorded onto a separate laptop computer.

In addition to the Kinects, to facilitate transcription, each
participant is wearing a set of Soundman OKM II Classic Studio binaural microphones. 
The audio from these is recorded via a Soundman A3 adapter onto Tascam DR-05 stereo recorders being worn by the participants. 
The recordings have been divided into training, development test, and evaluation test sets. 
Each set features non-overlapping homes and speakers. For more details about these datasets, see \cite{Fifth-Barker2018}.

\subsection{Tracks}
\label{tracks}
For the first time, the challenge moves beyond automatic speech
recognition (ASR) and also considers the task of diarization, i.e., estimating the start and end times and the speaker label of each utterance. The challenge features two tracks:
\begin{enumerate}
\item ASR only: recognise a given evaluation utterance given ground truth
  diarization information,
\item diarization+ASR: perform both diarization and ASR
\end{enumerate}
Both tracks are multi-array, i.e., all microphones of all arrays can be used. Track 1 is a rerun of the CHiME-5 challenge \cite{Fifth-Barker2018} and
Track 2 is similar to the ``Diarization from multichannel audio using system SAD'' track of the DIHARD
II challenge \cite{ryant2019second}, with the following key differences:
\begin{itemize}
\item an \textbf{accurate array synchronization} script is provided,
\item the \textbf{impact of diarization error on speech recognition error}
  will be measured,
\item \textbf{upgraded, state-of-the-art baselines} are provided for
  diarization, enhancement, and recognition.
\end{itemize}
These baselines and related implementations are integrated in the Kaldi speech recognition toolkit \cite{Kaldi-Povey2011} as a recipe.

For each track, we will produce two separate ASR rankings:
\begin{itemize}
\item[\textbf{A}] Systems based on conventional acoustic modeling and official language modeling: the outputs of the acoustic model must remain frame-level tied phonetic (senone) targets and the lexicon and language model must not be changed compared to the conventional ASR baseline,
\item[\textbf{B}] All other systems, including systems based on the end-to-end ASR baseline or systems whose lexicon and/or language model have been modified.
\end{itemize}
Ranking \textbf{A} focuses on acoustic robustness only,
while ranking \textbf{B} addresses all aspects of the scenario.

\section{Track 1}
\label{sec:track1_software}
Concerning Track 1, we provide baseline systems for array synchronization, speech enhancement, and speech recognition. 
All systems are integrated in the Kaldi CHiME-6 recipe\footnote{\url{https://github.com/kaldi-asr/kaldi/tree/master/egs/chime6/s5_track1}}.

\subsection{Overview}\label{sec:track1_software_overview}

The main script (\texttt{run.sh}) executes array synchronization, data preparation, data augmentation, feature extraction, Gaussian mixture model - hidden Markov model (GMM-HMM) training, data cleaning, and chain model training. 
After training, \texttt{run.sh} calls the inference script (\texttt{local/decode.sh}), which includes speech enhancement and recognition given the trained model. 
Participants can also execute \texttt{local/decode.sh} independently with their own ASR models or pre-trained models downloaded from the Kaldi model storage site\footnote{\url{http://kaldi-asr.org/models/m12}}.
Detailed technical descriptions of system components can be found in \cite{acoustic-manohar2019}. We outline the process below.

\begin{enumerate}
\item Array synchronization (stage 0)\\
  This stage first downloads the array synchronization tool, and generates the synchronized audio files
  across arrays along with their corresponding JSON files. Note that this requires
  sox v14.4.2, which is installed via miniconda in
  \texttt{./local/check\_tools.sh}. Details of the array
  synchronization procedure are presented in Section~\ref{sec:array-synchronization}.
\item  Data, dictionary, and language model (stages 1--3)\\
  These stages prepare data directories, the lexicon, and language models in the format expected by Kaldi. The lexicon has a 127,712 word vocabulary.
  We use a maximum entropy-based 3-gram language model, which achieves the best perplexity on the development set.
\item  Data augmentation (stages 4--7)\\
  In these stages, we augment and fix the training data. 
  Point source noises are extracted from the noise regions in the CHiME-6 corpus. 
  Here, we use a subset of 400~k utterances from the array microphones, their augmentations, and all worn microphone utterances during training.
  We did not include enhanced speech data for training to maintain the simplicity of the system.
\item  Feature extraction (stage 8)\\
  We extract 13-dimensional Mel-frequency cepstral coefficient (MFCC) features for GMM-HMM systems.
\item  GMM training (stages 9--13)\\
  These stages train monophone and triphone GMM-HMM models. These models are used for cleaning training data and generating lattices for training the
  chain model.
\item  Data cleaning (stage 14)\\
  This stage performs cleanup of the training data using the GMM model.
\item  Chain model training (stage 15)\\
  We use a factorized time delay neural network (TDNN-F) adapted from the Switchboard recipe 7q model \cite{semiorthogonal-povey2018}.
\item  Decoding (stage 16)\\
  This stage performs speech enhancement and recognition for the test
  set. This stage calls \texttt{local/decode.sh}, which includes
  speech enhancement (described in Section~\ref{sec:speech-enhancement}) and
  decoding and scoring (described in Section~\ref{sec:decoding-and-scoring}).
\end{enumerate}

\subsection{Array synchronization}\label{sec:array-synchronization}

The new array synchronisation baseline is available on GitHub\footnote{\url{https://github.com/chimechallenge/chime6-synchronisation}}. 
It compensates for two separate issues: \emph{audio frame-dropping} (which affects the Kinect devices only) and \emph{clock-drift} (which affects all devices).
It operates in the following two stages:

\begin{enumerate}
\item
  \emph{Frame-dropping} is compensated by inserting 0's into the signals where samples have been dropped. 
  These locations have been detected by comparing the Kinect audio with an uncorrupted stereo audio signal recovered from the video AVI files that were recorded (but not made publicly available). 
  The frame-drop locations have been precomputed and stored in the file \texttt{chime6\_audio\_edits.json}, which is then
  used to drive the synchronisation software.
\item
  \emph{Clock-drift} is computed by comparing each device's signal to the session's `reference' binaural recordings (the binaural mic of the speaker with the lowest ID number).
  Specifically, cross-correlation is used to estimate delays between the device and the reference at regular intervals throughout the recording session. 
  A relative speed-up or slow-down can then be approximated using a linear fit through these estimates. 
  The signal is then synchronised to the reference using a \texttt{sox} command to adjust the speed of the signal appropriately. 
  This adjustment is typically very subtle, i.e., less than 100~ms over a 2.5~h recording session. 
  Note, the approach failed for devices \emph{S01\_U02} and \emph{S01\_U05} which appear to have temporarily changed speeds during the recording session and have required a piece-wise linear fit. 
  The adjustments for clock-drift compensation have been precomputed and the parameters to drive the \texttt{sox} commands are stored in \texttt{chime6\_audio\_edits.json}.
\end{enumerate}

Note, after frame-drop and clock-drift compensation, the WAV files that are generated for each device will have slightly different durations.
For each session, device signals can be safely truncated to the duration of the shortest signal across devices, but this step is not performed by the synchronisation tool.

Finally, the CHiME-5 transcript JSON files are processed to fit the new alignment. 
In the new version, utterances will have the \textit{same} start and end time on every device.

\subsection{Speech enhancement}\label{sec:speech-enhancement}
We provide two baseline speech enhancement front-ends based on open-source implementations of guided source separation (GSS) \cite{boeddeker2018front} and BeamformIt \cite{Acoustic-Anguera2007}, respectively.
Both of them are combined with an open source version \cite{NaraWPE-Drude2018} of weighted prediction error (WPE) dereverberation \cite{nakatani2010speech}, and integrated into our Kaldi recipe. 
They can be installed in the Kaldi tool installation directory.

The first front-end consists of WPE, a spatial mixture model that uses time annotations (GSS), beamforming. 
They are applied to multiple arrays.
GSS is performed with the setup of \texttt{multiarray=outer\_array\_mics} meaning that only the first and last microphones of each array are used. 
This is the default speech enhancement front-end for the CHiME-6 Track 1 recipe.

The alternative front-end applies WPE based dereverberation and weighted delay-and-sum beamforming (BeamformIt) to the reference array. Users can easily switch from GSS to BeamformIt by specifying the enhancement option (e.g., \texttt{--enhancement beamformit}).

\subsection{Decoding and scoring}\label{sec:decoding-and-scoring}
We perform two-stage decoding, which refines i-vector extraction based on the first pass decoding result to achieve robust decoding of noisy speech \cite{acoustic-manohar2019}.
We also provide a scoring script for both development and evaluation: \texttt{local/} \texttt{score\_for\_submit.sh}.
The language model weight and insertion penalty are optimized based on the development set.

Note that, during scoring, we filter the tags ({[}noise{]}, {[}inaudible{]}, {[}laughs{]}, and {[}redacted{]}), and normalize ambiguous filler words\footnote{For example, we perform the following replacements to filter out variants of the filler word `hmm': \texttt{sed -e  's/\textbackslash<mhm\textbackslash>/hmm/g; s/\textbackslash<mm\textbackslash>/hmm/g; s/\textbackslash<mmm\textbackslash>/hmm/g;'}. The actual filtering rules can be found in \texttt{local/wer\_output\_filter}.}.

\section{Track 2}
\label{sec:track2_software}
Concerning Track 2, we provide baseline systems for array synchronization, speech
enhancement, speech activity detection (SAD), speaker diarization, and speech recognition. 
All systems are integrated in the Kaldi CHiME-6 recipe\footnote{\url{https://github.com/kaldi-asr/kaldi/tree/master/egs/chime6/s5_track2}}.

\subsection{Overview}\label{overview}
The main script (\texttt{run.sh}) is similar to \texttt{run.sh} in Track 1 as described in Section \ref{sec:track1_software}, which performs array synchronization, data preparation, data augmentation, feature extraction, GMM-HMM training, data cleaning, and chain model training. 
\texttt{run.sh} in Track 2 additionally includes SAD model training on the CHiME-6 dataset, and diarization model training on the VoxCeleb dataset \cite{nagrani2020voxceleb}.
We allow the participants to use VoxCeleb in addition to CHiME-6 data, since it is necessary to build a good diarization system.

After training, \texttt{run.sh} finally calls the inference script (\texttt{local/decode.sh}), which performs speech enhancement, SAD, speaker diarization, and speech recognition based on the trained models.
Participants can also execute \texttt{local/} \texttt{decode.sh} independently with their own SAD, diarization, and ASR models or pre-trained models.

Stages 1--7 are the same as those of Track 1, as described in Section~\ref{sec:track1_software_overview}. 
But these are followed by new subsequent stages:

\begin{enumerate}
\item SAD training\\
  We use a TDNN+LSTM (long short-term memory) model trained on the CHiME-6 dataset with ground truth alignments obtained by a GMM-HMM.
  Participants can also download a pretrained SAD model\footnote{\url{http://kaldi-asr.org/models/12/0012_sad_v1.tar.gz}}.
\item Diarization training\\
  An x-vector neural diarization model \cite{sell2018diarization} is trained with the VoxCeleb data \cite{nagrani2020voxceleb}. 
  This script is adapted from the Kaldi VoxCeleb v2 recipe.
  A probabilistic linear discriminant analysis (PLDA) model \cite{kenny2010bayesian} is trained on the CHiME-6 dataset.
  Participants can also download a pretrained diarization model\footnote{\url{http://kaldi-asr.org/models/12/0012_diarization_v1.tar.gz}}.
\item Decoding and scoring (stage 16)\\
    In Track 2, only raw recordings are given without segment or speaker information; i.e., \texttt{local/decode.sh} has to perform the whole pipeline consisting of speech enhancement, SAD, speaker diarization, and ASR decoding and scoring. These steps are detailed below.
\end{enumerate}

\subsection{Array synchronization}
Track 2 uses the exact same array synchronization technique as described in Section~\ref{sec:array-synchronization}.

\subsection{Speech enhancement}\label{speech-enhancement-track2}

Unlike Track 1, Track 2 only provides the BeamformIt-based
speech enhancement front-end (see Section~\ref{sec:speech-enhancement}) due to the risk of degradation in
GSS performance using estimated diarization information instead of ground truth speech segment information (which is unavailable).

\subsection{Speech activity detection}\label{sec:speech-activity-detection}
The SAD baseline relies on the neural architecture in \cite{ghahremani2016acoustic}.
It was trained using data (\texttt{train\_worn\_u400k}) from 1) the
CHiME-6 worn microphone utterances and 2) a randomly selected subset of 400~k array
microphone utterances.
We generate speech activity labels using an HMM-GMM system trained
with the \texttt{train\_worn\_simu\_u400k} data from 1) the CHiME-6
worn microphone utterances perturbed with various room impulse
responses generated from a room simulator and 2) a randomly selected subset of 400~k
array microphone utterances.

As a neural network architecture, we use 40-dimensional MFCC features as input, 5 TDNN layers, and 2 layers of statistics pooling \cite{ghahremani2016acoustic}.
The overall context of the network is set to be around 1~s, with around 0.8~s of left context and 0.2~s of right context.
The network is trained with a cross-entropy objective to predict speech/non-speech labels.

During inference, SAD labels for the test recordings are obtained by Viterbi decoding using an HMM with minimum duration constraints of 0.3~s for speech and 0.1~s for silence.
We also prepared an SAD decoding script to evaluate the SAD performance on the CHiME-6 data.
Note that the baseline system only performs SAD (and all other post-processing steps including speaker diarization and ASR) for the U06 array for simplicity. 
Exploring multi-array fusion techniques for SAD, diarization, and ASR is an integral part of the challenge.

\subsection{Speaker diarization}\label{sec:speaker-diarization}
The speaker diarization baseline relies on the segment files obtained by SAD. It is an x-vector system \cite{snyder2018x} with a 5-layer TDNN trained on the VoxCeleb dataset \cite{nagrani2020voxceleb}.
PLDA is trained on CHiME-6 data (\texttt{train\_worn\_simu\_u400k}).
Agglomerative hierarchical clustering (AHC) \cite{anguera2012speaker} is performed. Since the number of speakers in CHiME-6 is four in every session, this prior information is used by AHC.

Our speaker diarization system consistently uses the reference RTTM converted from the original
JSON file via data preparation (\texttt{run.sh\ -\/-stage\ 1}) by using the Kaldi RTTM conversion script\footnote{\url{https://github.com/kaldi-asr/kaldi/blob/master/egs/wsj/s5/steps/segmentation/convert_utt2spk_and_segments_to_rttm.py}}.
The diarization result is also obtained as an RTTM file, and the diarization error rate (DER) and Jaccard error rate (JER) are computed using dscore\footnote{\url{https://github.com/nryant/dscore}} (used in the
DIHARD II challenge).

Similar to the SAD system, this baseline system only performs diarization for the U06 array for simplicity. 

\subsection{Decoding and scoring}
The RTTM files obtained by speaker diarization in Section~\ref{sec:speaker-diarization} are converted to the Kaldi data format. 
We perform two-stage decoding, which refines the i-vector extraction based on the first pass decoding result to achieve robust decoding for noisy speech \cite{acoustic-manohar2019}. Again, the baseline system only performs ASR for the U06 array for simplicity.

We provide a scoring script for both development and evaluation. 
The language model weight and insertion penalty are optimized based on the development set. 
Multispeaker scoring is performed to obtain the concatenated minimum-permutation word error rate (cpWER). 

The cpWER is computed as follows:
\begin{enumerate}
	\item Concatenate all utterances of each speaker for both reference and hypothesis files. 
	\item Compute the WER between the reference and all possible speaker permutations of the hypothesis. There are 24 such permutations. 
	\item Pick the lowest WER among them (this is assumed to be the best permutation).
\end{enumerate}	
cpWER is directly affected by the speaker diarization results.
In addition to the cpWER, which shows the error rate of entire recordings, we also report detailed errors per utterance by recovering the utterance information from the reference.

\subsection{RTTM refinement}
\label{sec:rttm_refine}
In the original CHiME-5 annotations, utterance boundaries are marked by human annotators, among other information, in an RTTM file. While these utterance boundaries are sufficient for training and testing ASR, there are utterances that include long pauses between words, making them an imperfect reference for diarization. To obtain a more precise diarization reference, we apply forced alignment between the transcripts and the cleaner binaural recordings. 

The acoustic model we use for forced alignment is the triphone GMM-HMM model trained in the baseline system (see Section~\ref{sec:track1_software_overview}). We use \texttt{steps/align\_si.sh} to align the worn (binaural) microphone recordings for both the development set and the evaluation set. The decoding beam size is 20 by default, but if this search fails we perform a second alignment with a beam size of 150. These alignment results are at the word level. In order to merge them into utterances, we identify all contiguous words separated by gaps of at most 300~ms of silence. Instances of {[}noise{]} also separate utterances. These alignments are then saved in the RTTM format as the refined reference for SAD and diarization.

\section{Baseline results}
\label{sec:results}
\subsection{Track 1}
Table \ref{tab:track1baselineAsr} presents the Track 1 baseline ASR results with the official 3-gram language model (corresponding to category A). 
The CHiME-6 baseline is significantly better than the baseline and close to the best system \cite{du2018ustc} for the CHiME-5 multiple device track, which is equivalent to CHiME-6 Track 1.
Note that the CHiME-6 baseline is designed to be a compromise between performance and simplicity, e.g., we purposefully did not use system combination but the result is still close to \cite{du2018ustc}, which is based on complex multi-path enhancement processing with system combination.

\begin{table}[tbp]
	\centering
	\begin{tabular}{lcc}
		\toprule
		& Dev. WER & Eval. WER \\
		\midrule
		CHiME-6 baseline & 51.8\% & 51.3\% \\
		\midrule
		CHiME-5 baseline \cite{Fifth-Barker2018} & 81.1\% &	73.3\%  \\
		CHiME-5 top system \cite{du2018ustc} & 45.6\% & 46.6\%  \\
		\bottomrule
	\end{tabular}
	\caption{CHiME-6 Track 1 baseline ASR results, compared to the baseline and top systems for the (equivalent) CHiME-5 multiple device track.}
	\label{tab:track1baselineAsr}
\end{table}

\begin{table*}[tbp]
	\centering
	\begin{tabular}{l|ccc|ccc}
		\toprule
		& \multicolumn{3}{c}{Dev.}                  & \multicolumn{3}{c}{Eval.}                 \\
		& Missed speech & False alarm & Total error & Missed speech & False alarm & Total error \\
		\midrule
		Annotation RTTM & 2.5\%         & 0.8\%       & 3.3\%       & 4.1\%         & 1.8\%       & 5.9\%       \\
		Alignment RTTM & 1.9\%         & 0.7\%       & 2.6\%       & 4.3\%         & 1.5\%       & 5.8\%      \\
		\bottomrule   
	\end{tabular}
	\caption{CHiME-6 Track 2 baseline SAD results. Annotation RTTM is the original RTTM obtained by human annotation while alignment RTTM is based on forced alignment and is considered as the official RTTM file for the challenge.}
	\label{table:sad}
\end{table*}

\subsection{Track 2}
Table \ref{table:sad} shows the SAD performance obtained by the CHiME-6 baseline SAD system, as described in Section~\ref{sec:speech-activity-detection}.
It lists the SAD results computed against both human-annotated (old) and force-aligned (new) RTTMs, as described in Section~\ref{sec:rttm_refine}.
This result shows that the evaluation set is more difficult than the development set in terms of SAD performance.
Also, the new and old RTTMs have significant differences in the development set while they have marginal difference in the evaluation set.

\begin{table}[tbh]
	\begin{tabular}{l|cc|cc}
		\toprule
		& \multicolumn{2}{c}{Dev.} & \multicolumn{2}{c}{Eval.} \\
		& DER        & JER        & DER         & JER        \\
		\midrule
		Annotation RTTM & 61.6\%     & 69.8\%     & 62.0\%      & 71.4\%     \\
		Alignment RTTM & 63.4\%     & 70.8\%     & 68.2\%      & 72.5\%     \\
		\bottomrule   
	\end{tabular}
	\caption{CHiME-6 Track 2 baseline diarization results. Annotation RTTM is the original RTTM obtained by human annotation while alignment RTTM is based on forced alignment and is considered as the official RTTM file for the challenge.}
	\label{tab:track2baselineDiarization}
\end{table}

Table \ref{tab:track2baselineDiarization} shows the DER and JER obtained by the CHiME-6 baseline speaker diarization system, as described in Section \ref{sec:speaker-diarization}.
In spite of using a state-of-the-art diarization technique \cite{sell2018diarization}, both metrics show over 60\% error rates and improving the diarization performance is one of the main challenges in Track 2.

\begin{table}
	\centering
	\begin{tabular}{l|c|c|cc}
		\toprule
		&  & & Dev. & Eval. \\
		& Enhancement & Segmentation & WER & WER \\
		\midrule
		Track 1 & BeamformIt & Oracle & 69.8\% & 61.2\% \\
		Track 1 & GSS & Oracle & 51.8\% & 51.3\% \\
		Track 2 & BeamformIt & Diarization & 84.3\% & 77.9\% \\
		\bottomrule
	\end{tabular}
	\caption{CHiME-6 Track 1 and 2 baseline ASR results with BeamformIt-based \cite{Acoustic-Anguera2007} and GSS-based \cite{boeddeker2018front} speech enhancement. We used the same acoustic and language models for both tracks.}
\label{tab:track2baselineAsr}
\end{table}

Finally, Table \ref{tab:track2baselineAsr} shows the performance gap between the Track 1 and Track 2 baselines.
The main differences between the Track 1 and 2 baselines come from the use of advanced speech enhancement (GSS \cite{boeddeker2018front}, as described in Section~\ref{sec:speech-enhancement}) and the use of speech segmentation from manual annotations or automatic speaker diarization.
Note that both tracks use the same acoustic and language models.
Therefore, by comparing Tracks 1 and 2 with BeamformIt, we can observe that the main degradation (around 15\% absolute) comes from speaker diarization.

\section{Summary}
\label{sec:summary}
This paper describes the CHiME-6 challenge outline, baselines, and experimental results.
Newly introduced audio synchronization and a state-of-the-art Kaldi baseline simplify challenge entry for Track 1, while Track 2 significantly increases the difficulty due to the need for speaker diarization.
To help the challenge participants tackle these difficulties, we provide a complete set of open source Kaldi recipes for both Track 1 and Track 2 which combine speech enhancement, speaker diarization, and speech recognition.
This is the first trial in the community to provide open source recipes for unsegmented multispeaker ASR, and a lot of effort has been provided through volunteer activities by speech separation and recognition researchers in addition to the challenge organizers.
Our future work is to provide the analysis of this challenge including the investigation of the effectiveness of the techniques proposed during the challenge, a review of evaluation metrics, the relationship between diarization and speech recognition errors, and so on.
Through this analysis, we would like to make significant progress toward this challenging, realistic, and unsolved multispeaker speech processing problem.

\bibliographystyle{IEEEtran}
\balance
\bibliography{strings,refs,chime5}

\end{document}